\documentclass[aps,showpacs,floatfix,twocolumn,byrevtex,superscriptaddress]{revtex4-1}

\usepackage{amsmath}
\usepackage{amssymb}
\usepackage{amstext}
\usepackage{amsopn}
\usepackage{amsfonts}
\usepackage{amsxtra}
\usepackage[multiple]{footmisc}
\usepackage{bookmark}
\usepackage[english]{babel}
\usepackage{amsmath}
\usepackage{graphicx}
\usepackage{float}
\usepackage{bm}
\usepackage{multirow}
\usepackage{dcolumn}
\usepackage{hyperref}
\begin{document}


\title{Optical evidence of the band reconstruction during the charge-density wave transition in annealed Kagome magnet FeGe}

\author{A. Zhang}
\thanks{These authors contributed equally to this work.}
\affiliation{Key laboratory of Quantum Materials and Devices of Ministry of Education, Department of Physics, Southeast University, Nanjing 211189, China}
\author{X. -L. Wu}
\thanks{These authors contributed equally to this work.}
\author{R. Yang}
\thanks{These authors contributed equally to this work.}
\email{ryang@seu.edu.cn}
\affiliation{Key laboratory of Quantum Materials and Devices of Ministry of Education, Department of Physics, Southeast University, Nanjing 211189, China}
\author{A. -F. Wang}
\affiliation{Low Temperature Physics Laboratory, College of Physics and Center of Quantum Materials and Devices, Chongqing University, Chongqing 401331, China}
\author{Y. -M. Dai}
\email{ymdai@nju.edu.cn}
\affiliation{National Laboratory of Solid State Microstructures and Department of Physics, Nanjing University, Nanjing 210093, China}
\author{Z. -X. Shi}
\email{zxshi@seu.edu.cn}
\affiliation{Key laboratory of Quantum Materials and Devices of Ministry of Education, Department of Physics, Southeast University, Nanjing 211189, China}
\date{\today}

%

\begin{abstract}
In Kagome magnet FeGe, the coexistence of electron correlation, charge-density wave (CDW), and magnetism renders it ideal to study their interactions.
Here, we combined the optical spectroscopy and the first-principles calculations to investigate the band structures of FeGe annealed at different temperatures.
Our observations reveal that the sample annealed at 320$^\circ C$ experienced dramatic change in optical conductivity following the CDW transition. Specifically, a substantial portion of the spectral weight (SW) in the low-energy region ($<$ 0.4~eV) was redistributed to the high-energy region (0.8 $\sim$ 1.5~eV), suggesting a reconstruction of the band structure.
The sample annealed at 560$^\circ C$ did not exhibit a CDW transition, but its SW transfer occurred progressively from 300 to 5~K.
We noticed that: i) after the CDW transition, the sample annealed at 320$^\circ C$ showed similar tendency of SW transfer to that of the 560$^\circ C$ annealed sample; ii) the high-energy SW of both materials displayed a temperature dependence consistent with the magnetic properties. Combining the first-principles calculations, we attribute the SW transfer to the band reconstruction triggered by the distortion of Ge1 atoms induced either by annealing at 560$^\circ C$ or by the CDW transitions. This lattice distortion affects the energies of Fe $3d$ orbitals. Under the influence of Hund's rule coupling, the magnetic moment of Fe atoms is enhanced. Our findings elucidate the interactions among charge, lattice, and spin in FeGe, offering pivotal insights to modulate properties of this Kagome magnet.
\end{abstract}


\maketitle
%
\section{Introduction}
Transition metal compounds exhibit a diverse array of physical properties due to the intricate distribution of electrons within their $d$ orbitals, which is influenced by factors such as lattice structure, strong electron correlation, and the Hund's rule coupling.
This complexity leads to phenomena such as high-temperature superconductivity observed in cuprates~\cite{yu2019high,can2021high,mankowsky2014nonlinear} and iron-based materials~\cite{fernandes2014drives, sun2019review, Stewart2011ironreview}, colossal magnetoresistance in manganese oxides~\cite{subramanian1996colossal}, and orbital selectivity in ruthenium oxides~\cite{khalifah2002orbital}.
These phenomena significantly enriched the condensed matter physics, and the interplay of charge, spin, and lattice degrees of freedom offers a promising avenue for quantum modulation~\cite{tokura2000orbital}.
In recent years, transition metal compounds with Kagome lattices have stimulated a wide range of interest.
There has been a surge of interest in transition metal compounds featuring Kagome lattices, which are known for their unique structure and the novel physical phenomena~\cite{liu2024superconductivity,yin2018giant,jiang2023kagome}.
In insulating magnets, where localized moments predominate, the geometric frustration inherent to the Kagome lattice can give rise to quantum spin liquid states that do not exhibit spontaneous symmetry breaking~\cite{yin2018giant, Broholm2020}.
In metallic systems, the destructive interference of electronic wavefunctions within the Kagome lattice can simultaneously result in the formation of flat bands, topological states, and van Hove singularities (VHS)~\cite{yi2024polarizedchargedynamicsnovel, WangY2023, chen2024competing}.
FeGe, a material that combines magnetism, itinerant carriers, and a Kagome lattice, has emerged as an exemplary platform for investigating the synergistic effects of these three elements~\cite{WangY2023, teng2022discovery, Wushangfei2024PRX}.

\begin{figure*}[t]
\centerline{
\includegraphics[width=2\columnwidth]{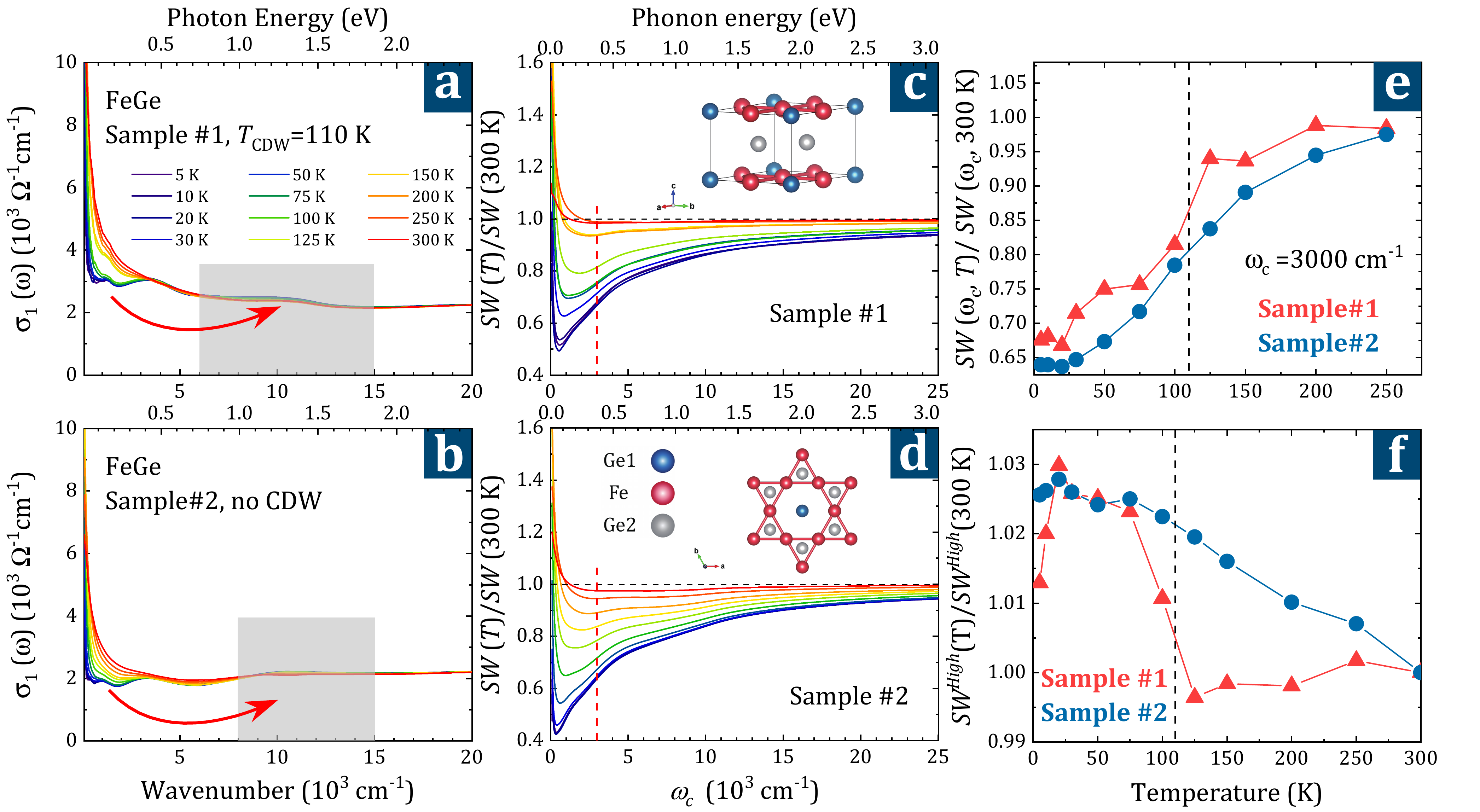}
}
\caption{Optical spectroscopy of FeGe samples annealed at different temperatures. (a)-(b) Temperature evolution of the real part $\sigma_1(\omega;T)$ of the optical conductivity spectra of (a) Sample $\#1$ (S1) and (b) Sample $\#2$  (S2) below 25\,000~cm$^{-1}$(3.1~eV); the tendency of $SW$ shifting in $\sigma_1(\omega;T)$ is indicated by the red arrows. (c)-(d) The integrated spectral weight ratio $SW(T) / SW(300 K)$ as a function of cutoff frequency ($\omega_c$) at different temperatures. $S W\left(\omega_c ; T\right)=\left(Z_0 / \pi^2\right) \int_0^{\omega_c} \sigma_1\left(\omega^{\prime} ; T\right) d \omega^{\prime}$, in which $Z_0=377~\Omega^{-1}$ is the vacuum impedance. The insets display the lattice structure of FeGe in different orientations, which contain two inequivalent Ge atoms (i.e. Ge1 and Ge2 atoms). (e) and (f) display the spectral weight normalized by that at 300~K, $SW(\omega_c,T) / SW(\omega_c,300 K)$, for cut off frequency $\omega_c$ = 3000~cm$^{-1}$ and $\omega_c$ = 8000~cm$^{-1}$, respectively.}
\label{fig:optical spectroscopy}
\end{figure*}

The B35 phase of FeGe is characterized by a Kagome lattice composed of iron (Fe) atoms, with two distinct germanium (Ge) atoms positioned within this lattice: Ge1 atoms are situated in the Kagome layer, while Ge2 atoms form intercalated honeycomb layers(as shown in the insets of Figs~\ref{fig:optical spectroscopy}c and d).
Below the N$\acute{e}$el temperature ($T_N\sim$ 410 K), an A-type antiferromagnetic (AFM) order is established among the Fe atoms.
As the temperature ($T$) drops below 110~K, a lattice distortion occurs, resulting in a 2$\times$2$\times$2 charge modulation~\cite{chen2024discovery,wang2023enhanced,zhao2023photoemission,yin2022discovery}.
Concurrent with the CDW transition, there is a significant enhancement in the magnetic moment of the Fe atoms,suggesting a complex interplay between the lattice, spin, and charge degrees of freedom within FeGe.
Recent studies revealed very close energies of the three degrees of freedom of lattice, charge, and spin in FeGe, and the occurrence of charge ordering and lattice distortion is intended to save magnetic properties~\cite{wang2023enhanced, Teng2024}.
Nevertheless, the precise nature of the coupling between spin, charge, and lattice in FeGe remains to be elucidated.

%
%

Recent studies have underscored the pivotal influence of the distortion of Ge1 atoms along the c-axis on the physical properties of FeGe~\cite{yi2024polarizedchargedynamicsnovel,wenzel2024intriguing, Chen2024, Han2024FeGeorbital}.
Wu \emph{et al.} have demonstrated that by manipulating the annealing temperature, the distortion of Ge1 sites in as-grown samples can be effectively regulated, thereby modulating the CDW transition and magnetic properties.~\cite{wu2024annealing}.
In samples subjected to annealing at ${320}^{\circ} \mathrm{C}$, Ge1 atoms locate within the Fe$_3$Ge layers and display long-range CDW order below 110~K.
Conversely, in samples annealed at ${560}^{\circ} \mathrm{C}$, $8\%$ of the Ge1 atoms exhibited distortion relative to the Fe layer, which disrupts the CDW phase transition but elevates the N$\acute{e}$el temperature.
Due to the significant hybridization between the Ge1 $p$-orbitals and the $3d$-orbitals of Fe, the distortion of Ge1 atoms inevitably affects the bands embracing Fe~$3d$ orbitals.
Consequently, comparative analyses of samples annealed at distinct $T$s are instrumental in elucidating the mechanisms governing CDW phase transitions in FeGe, as well as the intricate interplay among charge, spin, and lattice dynamics preceding and succeeding the phase transition.

In this work, we conducted a comparative study on FeGe samples annealed at ${320}^{\circ} \mathrm{C}$ (Sample \#1, S1) and ${560}^{\circ} \mathrm{C}$ (Sample \#2, S2) using the optical spectroscopy.
We found that the distortions of Ge1 atoms along the c-axis caused either by the CDW transitions or by the annealing at ${560}^{\circ} \mathrm{C}$ can suppress the charge excitations below 0.4~eV in optical conductivity while enhancing the excitations in the 0.8 to 1.5~eV range, suggesting that lattice distortions result in a band reconstruction.
Combining first-principles calculations, we propose that highly degenerate VHS close to the Fermi level are responsible for the structural instability, and the distortion of Ge1 atoms leads to the band reconstruction, altering the energy of Fe $3d$ orbitals.
Under the influence of Hund's coupling, the charge and spin distribution of electrons within these orbitals is perturbed, leading to significant changes in electrical and magnetic properties of FeGe.

\section{Results}


\subsection{Optical spectroscopy}

We measured the $T$-dependent reflectivity of two samples (S1 and S2, Fig. S1 of the supplemental materials (SM)~\cite{Supplementary}) in a broad energy range (0.01$\sim$3~eV). Through the Kramers-Kronig transformation, we derived the real part of the optical conductivity($\sigma_1(\omega)$) that reflects the joint density of states of materials (as shown in Figs.~\ref{fig:optical spectroscopy}a and b, we refer to Sec. I of the SM~\cite{Supplementary} for detail of the measurements).
From the $\sigma_1(\omega)$s, it is evident that both materials exhibit distinct mentality characterized by Drude peaks at zero frequency attributable to the intraband responses, and high-frequency Lorentz peaks dominated by a series of interband transitions.
Upon cooling, $\sigma_1(\omega)$s of both samples show a pronounced $T$ dependence.
For sample S1 (Fig.~\ref{fig:optical spectroscopy}a), the low-frequency response ($<$ 0.4~eV) is significantly suppressed after the CDW transition at 110~K, while the high-frequency response (0.8$\sim$ 1.5~eV) gradually increases, indicating the band reconstruction induced by the CDW transition.
Conversely, for sample S2, despite the absence of the CDW transition, both the low-frequency and high-frequency optical conductivities exhibit continuous reduction and enhancement, respectively, from 300 to 5~K.
This behavior is reminiscent of the high-energy pseudogap observed in iron-based superconductors~\cite{WangNL2012pesudogap}.
It is worth mentioning that, comparing with S1, the $\sigma_1(\omega)$s of S2 shows a significant suppression below 0.4~eV and enhanced strength in the high-energy region (Fig. S2 of the SM~\cite{Supplementary}) at 300~K, indicating that, in S2, the band reconstruction already happens at high $T$ without CDW transition.

To reveal the correlation between the variations in low- and high-frequency $\sigma_1(\omega)$s, in Figs.~\ref{fig:optical spectroscopy}c and d, we plotted the ratio of the $SW$ at low $T$s over that at 300~K.
The results indicate that while the low-frequency $SW$ for both samples gradually decreases at low $T$s, it gradually recovers as the integration interval increases, suggesting a progressive transfer of the low-frequency $SW$ to the high-energy range, as indicated by the arrows in Figs.~\ref{fig:optical spectroscopy}a and b.
In Fig.~\ref{fig:optical spectroscopy}e, we calculated the $SW$ of $\sigma_1(\omega)$ below 3000~cm$^{-1}$ for both samples.
For sample S1, the low-frequency $SW$ exhibits no significant $T$ dependence above 110~K.
However, below 110~K, a substantial portion of the low-frequency $SW$ is abruptly suppressed, and diminishes progressively with further decreasing $T$.
In contrast, for sample S2, the suppression of the low-frequency $SW$ exists from 300 to 5~K.
We noticed that, after the lattice distortion in sample S1, the trend of $SW$ transfer in both samples becomes consistent.

On the other hand, high-frequency excitations are associated with localized orbitals~\cite{zhao2023photoemission, yang2024charge}.
In Fig.~\ref{fig:optical spectroscopy}f, we calculated the $SW$ of $\sigma_1(\omega)$ in the high-frequency range (the gray area in Figs.~\ref{fig:optical spectroscopy}a and b) for both samples.
Our observations reveal that the high-frequency $SW$ of sample S1 increased abruptly following the CDW transition but exhibited a slight decrease subsequent to the spin canting at 40~K.
In the case of sample S2, the high-frequency $SW$ gradually increased with decreasing $T$, and similarly displayed a sudden decrease after the spin canting at low $T$s.
Notably, the $T$-dependent behavior of the high-frequency $SW$ in both samples is highly consistent with the $T$ dependence of the magnetic moments of Fe atoms, as determined by neutron scattering~\cite{teng2022discovery}, and the magnetic susceptibility measurements~\cite{wu2024annealing}.
This suggests that high-energy excitations may be linked to magnetic properties in FeGe.

%
\begin{figure}[ht]
	\centerline{
		\includegraphics[width=1\columnwidth]{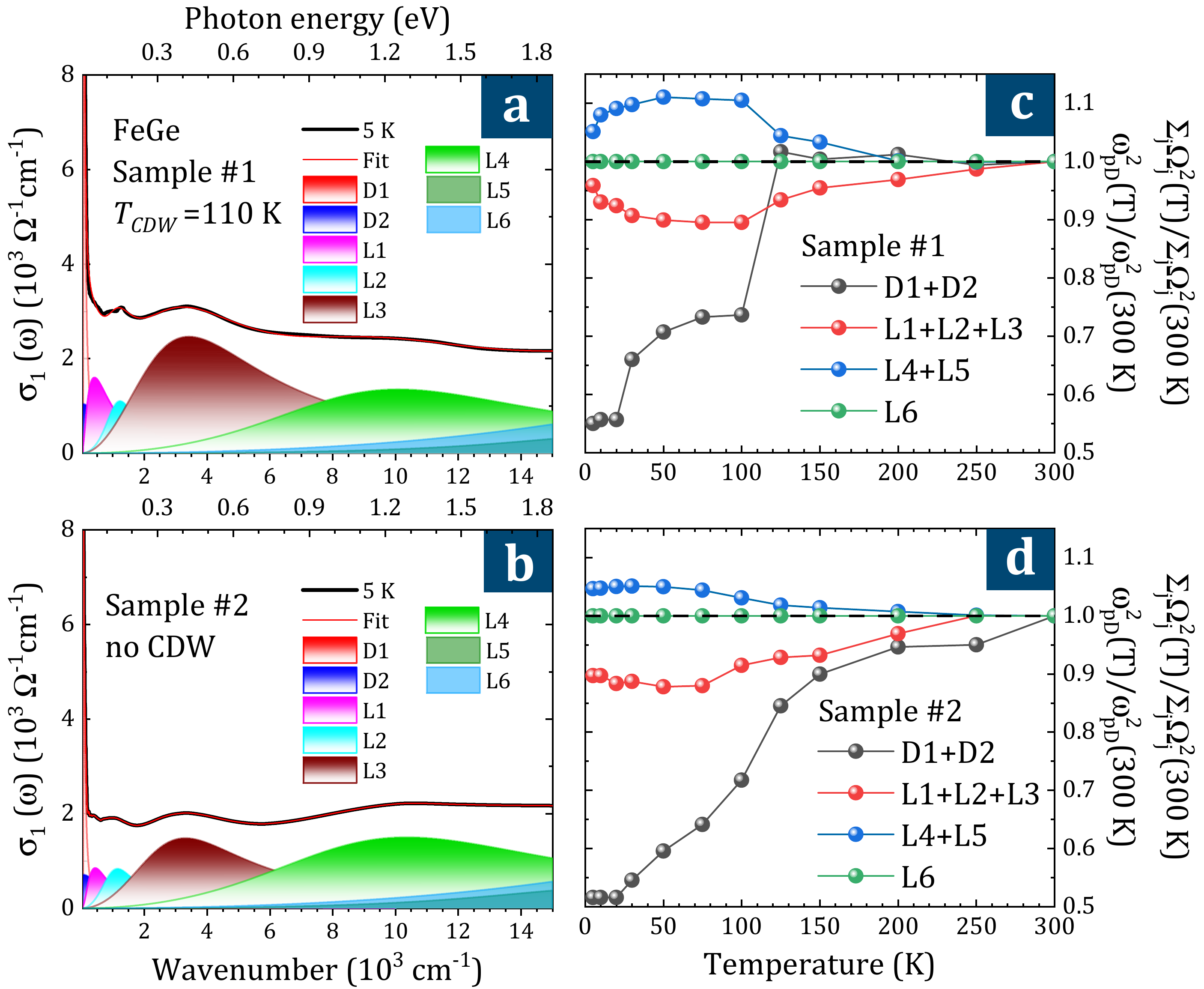}
	}
	\caption{Drude(D)-Lorentz(L) fit to the optical conductivity. (a) and (b) are the fits to optical conductivities at 5 K for two FeGe samples annealed at different temperatures; the narrow and wide Drude components, as well as several Lorentz oscillators, can be extracted from the optical conductivities. (c) and (d) display the spectral weights of Drude ($\omega_{p D}^2=\omega_{p D 1}^2+\omega_{p D 2}^2$) and Lorentz ($\sum_j \Omega^2$) components, normalized to the value at 300 K, respectively. We refer to Sec. II of the SM~\cite{Supplementary} for further information of the Drude-Lorentz model.}
	\label{fig:fit}
\end{figure}

%
\begin{figure*}[ht]
	\centerline{
		\includegraphics[width=2\columnwidth]{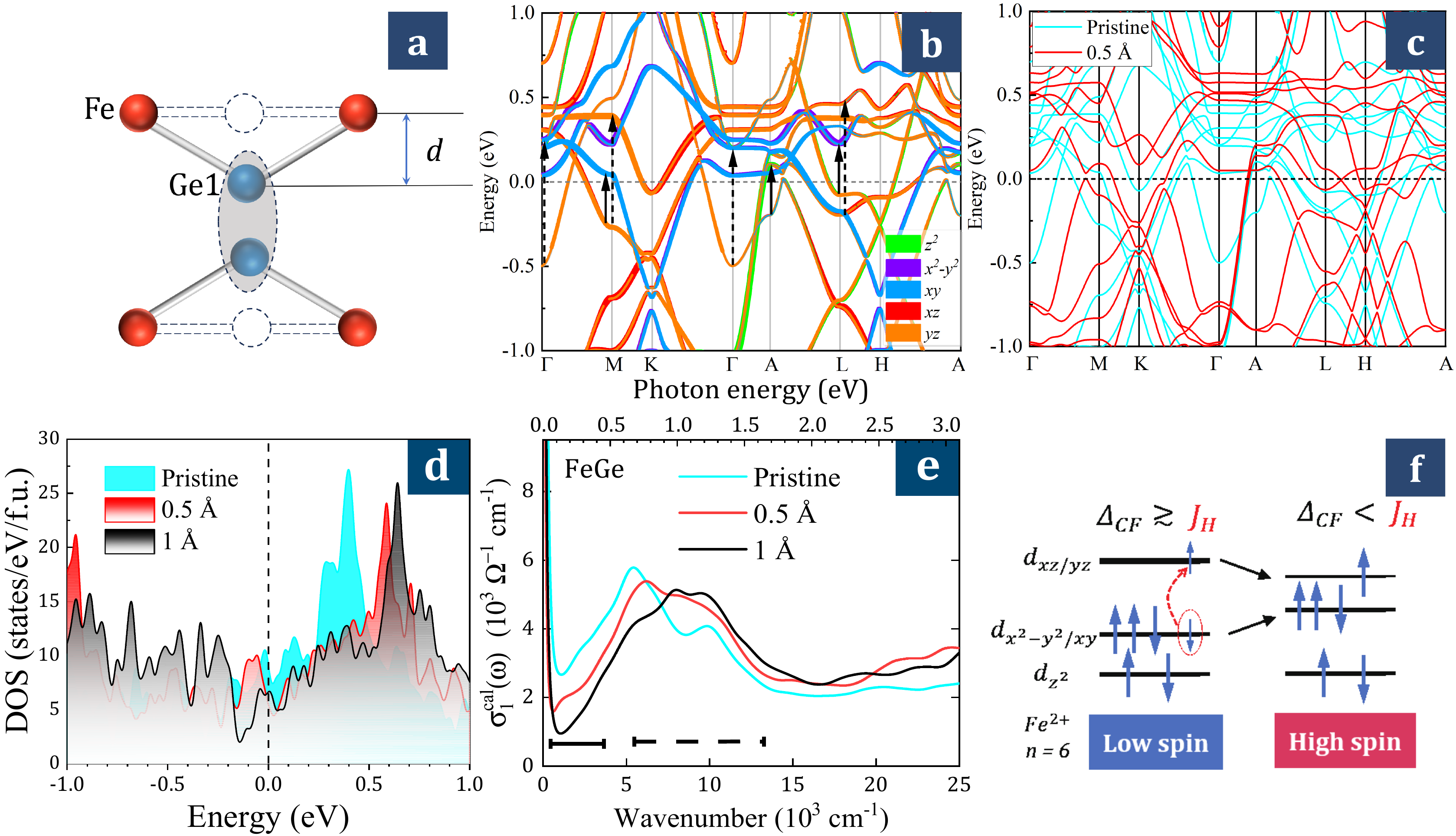}
	}
	\caption{The first-principles calculations. (a) is the sketch of dimer formation by Ge1 atoms. (b) displays the orbital-projected band structure of FeGe close to the Fermi level. The local coordinate is used to perform the Fe-$3d$ orbital projection. (c)-(e) The energy band structures, density of states, and optical conductivities of FeGe when Ge1 atoms undergo different degrees of distorted. (f) shows a schematic representation of the evolution of Fe $3d$ orbitals from the low spin state to the high spin state under the influence of crystal field splitting ($\Delta_{CF}$) and Hund's coupling ($J_H$). }
	\label{fig:DFT}
\end{figure*}

\subsection{Drude-Lorentz fit}
To quantitatively describe the electrodynamic response of these two samples, we fitted the $\sigma_1(\omega)$ based on the Drude-Lorentz mode (refer to Sec. II of the SM~\cite{Supplementary} for details).
As shown in Figs.~\ref{fig:fit}a and b, the $\sigma_1(\omega)$ for both samples can be resolved into two Drude (D) peaks, corresponding to intraband transitions, and a series of Lorentz (L) peaks, indicative of interband transitions.
This allows us to quantitatively analyze the $SW$ of each component.
With decreasing $T$, as depicted in Figs.~\ref{fig:fit}c and d, variations in the $SW$ of different components become apparent.
In sample S1, following the CDW transition at 110~K, the low-energy $SW$, encompassing both intraband responses (D1 and D2) and interband transitions (L1 to L3), undergoes suppression to varying degrees, while the $SW$ of higher-energy interband transitions (L4+L5) gradually increases, suggesting a redistribution of $SW$ from lower to higher frequencies.
Below 40~K, when sample S1 undergoes spin canting, the $SW$ of L4+L5 is subsequently transferred back to the lower-energy interband transitions (L1 $\sim$ 3).
Notably, the $SW$ evolution of L4+L5 also correlates with the trend in Fe's magnetic moments measured by neutron scattering~\cite{teng2022discovery}.
Conversely, in sample S2, as $T$ decreases from 300~K, a gradual and continuous transfer of $SW$ from low-energy components to higher frequencies was observed.
Although CDW transition is undiscernible in S2, the redistribution of $SW$ in S1 after the CDW transition resembles that of S2.
This implies that ${560}^{\circ} \mathrm{C}$ annealing and the CDW transition have analogous impacts on the lattice as well as the band structure.

\subsection{The first-principles calculations}
In sample S1, the most significant change after the CDW transition is the displacement of $1/4$ of the Ge1 atoms in the Fe$_3$Ge layer along the c-axis, forming a dimer with Ge1 atoms in the neighboring layer (as depicted in Fig.~\ref{fig:DFT}a).
In sample S2, despite the absence of a CDW phase transition, ${560}^{\circ} \mathrm{C}$ annealing has already resulted in the distortion of Ge1 atoms (around 8\%) along the c-axis.
Therefore, we speculate that the variations observed in the optical spectra of both samples may originate from the distortion of the Ge1 atoms.
To elucidate the impact of the distortion of Ge1 atom on the band structure, we calculated the band structure as well as the $\sigma_1(\omega)$ while taking into account lattice distortion.
Given that only a quarter of the Ge1 atoms in S1 are subject to the distortion, precise calculations necessitate the consideration of a large unit cell, which surpasses our computational resources.
Nevertheless, recent studies have indicated that before and after the CDW transition, only the Ge1 atoms exhibit significant distortion in the lattice, with minimal changes at other lattice sites, and only the bands related to the Ge1 atoms have undergone changes in the band structure~\cite{Zhang2024}.
Therefore, to qualitatively trace the impact of Ge1 atom distortion on the band structure, we only considered the distortion within one unit cell during the calculation.
Our results are qualitatively consistent with recent theoretical calculations that incorporated larger supercells and the angle-resolved photoemission spectroscopy (ARPES)  measurements~\cite{wang2023enhanced,zhao2023photoemission,shao2023intertwining, Zhang2024}.

Firstly, we calculated the band structure without the lattice distortion in Fig. 3b.
We found that the bands proximate to the Fermi level are predominantly constituted by the $3d$ orbitals of Fe atoms.
In certain bands, there is a noticeable overlap between the $3d$ orbitals of Fe and the $4p$ orbitals of Ge1, suggesting substantial hybridizations (the distribution of Ge1's $4p$ orbital is given in Fig. S5 of the SM~\cite{Supplementary}).
Therefore, the distortion of Ge1 atoms is anticipated to exert an influence on Fe $3d$ orbitals.
Due to the Kagome lattice formed by Fe atoms, their $3d$ orbitals give rise to flat bands positioned approximately 0.5~eV above the Fermi level, a Dirac cone along the $\Gamma-M$ direction, and a VHS 0.3~eV below the Fermi level at the $M$ point.

Next, in Fig.~\ref{fig:DFT}c, we calculated the band structure considering the distortion of the Ge1 atom along the c-axis (red) and compared it with the pristine band structure (blue) (for details on the evolution of the bands, refer to Fig. S3 in the SM~\cite{Supplementary}).
We found that, accompany with the lattice distortion, the flat bands gradually shift towards higher energies.
Meanwhile, for the occupied states, the Fe $d_{yz}$ orbital at the $\Gamma$ point gradually shifts downwards, increasing the occupancy.
Conversely, at the $K$ and $H$ points, bands contributed by $d_{xy}$/$d_{x^2-y^2}$ orbitals shift upwards, resulting in a decrease in the occupied states.
It is noteworthy that, as the Ge1 atom distorts, the VHS at the $M$ point is getting closer to the Fermi level.
The VHS at the $M$ point, formed by the degeneracy of $d_{xz}$ and $d_{yz}$ orbitals, possesses a very high density of states, which can easily lead to the instability of the crystal structure.

To further trace the impact of Ge1 atom distortion on $\sigma_1(\omega)$, based on the band structure, we calculated the $\sigma_1(\omega)$s of FeGe with different heights of the Ge1 atoms.
We found that, as the Ge1 atom deviates from the iron plane, intraband and low-energy excitations are progressively diminished, leading to a shift of the $SW$ towards higher-energy excitations.
Considering the energy scale and the spin polarizations (refer to Fig. S4 of the SM~\cite{Supplementary} for bands of different spin polarizations), we attributed the low-energy (solid segment) and high-energy (dashed segment) excitations in $\sigma_1(\omega)$ to direct interband transitions indicated by corresponding solid and dashed arrows in Fig.~\ref{fig:DFT}b.
The distortion of the Ge1 atom affects the band structure, gradually shifting the interband transitions to higher energies.
Concomitantly, the density of states at the Fermi level decreases after the distortion of the Ge1 atom (Fig.~\ref{fig:DFT}d), corresponding to the observed suppression of intraband (Drude) responses.

\section{Discussion}

Although only a fraction of Ge1 atoms experience distortion, the consistency between observations and theoretical calculations confirmed that the changes in $\sigma_1(\omega)$ of both samples mainly come from the band reconstruction caused by the distortion of Ge1 atoms.
Subsequently, it is imperative to elucidate the factors precipitating the lattice distortion and to investigate the ramifications of Ge1 atom distortion on magnetic properties.

From band structure calculations, we find that the Kagome lattice formed by Fe atoms results in flat bands and VHS, both of which significantly contribute to the density of states.
The VHS at the $M$ point, which is a degeneracy for the $d_{xz}$/$d_{yz}$ orbitals, is very close to the Fermi level (it will be more closer when electron correlations are considered~\cite{wang2023enhanced}).
In real space, the high electron density in the $d_{xz}$/$d_{yz}$ orbitals leads to lattice instability along the c-axis.
Thus, a minor displacement of Ge1 atoms along the c-axis during the cooling process can bring the VHS closer to the Fermi surface~\cite{wu2024annealing}.
This shift can induce long-range lattice instability, culminating in the lattice distortion that is observed in sample S1~\cite{Zhang2024}.

While the $3d$ orbitals of Fe are all located near the Fermi level with fairly close energies, the crystal symmetry of FeGe results in a discernible energy difference among these orbitals.
They can be categorized into three groups: $d_{z^2}$, $d_{xz}$/$d_{yz}$, and $d_{xy}$/$d_{x^2-y^2}$~\cite{wu2023electron, Huang2020}.
When Ge1 atoms are in the iron plane, the Ge1 $p_z$ orbitals overlap significantly with the Fe $d_{xz}$/$d_{yz}$ orbitals, resulting in a higher energy for these orbitals compared to $d_{x^2-y^2}$/$d_{xy}$ orbitals.
This energy difference slightly exceeds the Hund's coupling energy ($J_H\sim$0.8~eV)~\cite{Huang2020}.
In addition, the pronounced hybridization between Ge1 $p$ and Fe 3$d$ orbitals endows carriers with considerable kinetic energies, consequently limiting the number that can be bound and polarized onto the $d_{xz}$/$d_{yz}$ orbitals via Hund's coupling, resulting in a low-spin state of Fe$^{2+}$ (Fig.~\ref{fig:DFT}f)~\cite{Wu2008orbital}.
When Ge1 atoms distort along the c-axis, the hybridization between their $p$-orbitals and the Fe $d$-orbitals diminishes, lowering the energy of the $d_{yz}$ orbital and raising that of the $d_{x^2-y^2}$/$d_{xy}$ orbitals (as shown in the band structure in Fig.~\ref{fig:DFT}c), and narrowing the energy difference between these two sets of orbitals to below the $J_H$.
Concurrently, the diminished hybridization reduces the kinetic energy of the carriers (flattening the bands), prompting more electrons to be bound and  polarized onto the $d_{xz}$/$d_{yz}$ orbitals and giving rise to a high-spin state(Fig.~\ref{fig:DFT}f)~\cite{Huang2020, WangNL2012pesudogap}.
In the calculated $\sigma_1(\omega)$s, an enhancement above 0.8~eV indicates the increasing occupation of the $d_{xz}$/$d_{yz}$ orbitals (Figs.~\ref{fig:DFT}b and c), aligning the high-energy $SW$ with the tendency of the magnetism.
At low $T$s, when spin canting happens in both samples, spin-orbit coupling effects lead to the redistribution of electrons among Fe $3d$ orbitals, slightly reducing the high-energy $SW$ and magnetism~\cite{Lou2024}.
Our calculations further demonstrate that elevating the Ge1 atoms gradually strengthens the magnetism at the Fe site (refer to Fig. S6 in SM~\cite{Supplementary} for details on the correlation between Ge1 atom distortion and the magnetic moment of Fe atom).

In sample S2, 8\% of Ge1 atoms have already distorted along the c-axis after the annealing at 560$^\circ C$ , thus its band structure has transformed above 300~K, transferring the low-energy $SW$ to the high-energy region (Fig. S2 of SM~\cite{Supplementary}).
Consequently, Fe atoms are driven into a high-spin state at much higher temperatures, which correlates with the observed increase in the $T_N$ for sample S2.
Upon cooling, the kinetic energy of the itinerant carriers diminishes, and the effect of Hund's coupling becomes more pronounced, progressively binding and polarizing the itinerant electrons.
This process is reflected by the continuous shift of the $SW$ in S2's $\sigma_1(\omega)$ from the low- to the high-energy region, aligning with the behavior observed in iron-based unconventional superconductors, in which an increase in the As/Se content is associated with enhanced Hund coupling effects~\cite{WangNL2012pesudogap, wang2023enhanced, Gerber2017FeSeorbital, Yin2011}.

\section{Conclusion}
In summary, we employed the optical spectroscopy measurements and the first-principles calculations to conduct a comparative investigation on FeGe single crystals annealed at different temperatures. Our findings revealed that during the CDW transition in FeGe, a distortion of Ge1 atoms along the c-axis triggers a significant band reconstruction. The lattice distortion changes the hybridization between Ge1 $p$ and Fe $3d$ orbitals, leading to an enhancement of the magnetic moment at the Fe sites under the influence of Hund's coupling. Moreover, we found that annealing at ${560}^{\circ} \mathrm{C}$ can also distort Ge1 atoms and modify the band structure in the same way but in much broader temperature range. Our research reveals the intricate interplay between lattice distortions, charge dynamics, and spin states in FeGe, offering crucial insights for the manipulation of its physical properties.

%
%
\section{Acknowledgments}
We thank B. Xu for fruitful discussions.
Y. -M. Dai acknowledges the support from the National Natural Science Foundation of China (Grant Nos. 12047503 and 12374165).
A. Wang acknowledges the support from the National Natural Science Foundation of China (No. 12474142 and No. 12004056)
R.Yang acknowledges the support from the open research fund of Key Laboratory of Quantum Materials and Devices (Southeast University) Ministry of Education, the Alexander von Humboldt foundation, and the National Natural Science Foundation of China (Grant No. 12474152).
Numerical computations were performed on Hefei advanced computing center and Big Data Computing Center of Southeast University.
The authors thank the Center for Fundamental and Interdisciplinary Sciences of Southeast University for the support in optical spectroscopy measurements.
%
%
%
%
%
%
\section{Author contribution}
X. -L. W. and A. -F. W. grew the single crystals and carried out the transport measurements. Y. -M. D. and R. Y carried out the optical measurements. A. Zhang performed the first-principles calculations. R. Y. analyzed the data and prepared the manuscript with comments from all authors. R. Y. and Z. -X. S. supervised this project.
%
%

\end{document}